\documentclass[]{aastex631}

\submitjournal{ApJ}
\shortauthors{Rahmanifard et al.}
\usepackage{multirow}

\begin{document}

\title{The Effect of Angular Scattering Imposed by Charge Exchange and Elastic Collisions on Interstellar Neutral Hydrogen Atoms}

\author[0000-0001-9316-0553]{F. Rahmanifard}
\affiliation{Physics Department, Space Science Center,\
University of New Hampshire,\
Durham, NH 03824, USA }

\author{P. Swaczyna}\altaffiliation{currently at Space Research Centre PAS (CBK PAN), Bartycka 18A, 00-716 Warsaw, Poland}
\affil{Department of Astrophysical Sciences,\
Princeton University,\
Princeton, NJ 08544, USA}

\author{E. J. Zirnstein}
\affil{Department of Astrophysical Sciences,\
Princeton University,\
Princeton, NJ 08544, USA}

\author{J. Heerikhuisen}
\affil{Department of Mathematics and Statistics,\
University of Waikato,\
Hamilton, New Zealand}

\author{A. Galli}
\affiliation{Physics Institute,\
University of Bern,\
Bern, 3012, Switzerland }

\author{J. M. Sok\'{o}\l}
\affil{Southwest Research Institute,\
San Antonio, TX 78228, USA}

\author{N. A. Schwadron}
\affiliation{Physics Department, Space Science Center,\
University of New Hampshire,\
Durham, NH 03824, USA }

\author{E. M\"{o}bius}
\affil{Physics Department, Space Science Center,\
University of New Hampshire,\
Durham, NH 03824, USA }

\author{D. J. McComas}
\affil{Department of Astrophysical Sciences,\
Princeton University,\
Princeton, NJ 08544, USA}

\author{S. A. Fuselier}
\affil{Southwest Research Institute,\
San Antonio, TX 78228, USA}
\affil{University of Texas at San Antonio,\
San Antonio, TX 78228, USA}



\begin{abstract}

Angular scattering in charge exchange and elastic collisions between interstellar ions and neutral (ISN) atoms has been assumed to be negligible in previous studies. Here, we investigated the momentum transfer associated with the angular scattering of H atoms using Monte Carlo calculations to simulate their transport through the outer heliosheath. We considered two cases where charge exchange and elastic collisions between ISN H atoms and protons occur with and without momentum transfer in the outer heliosheath. We then simulated the transport of ISN H atoms inside the heliosphere to simulate count rates observed in the lowest energy bin of IBEX-Lo. We studied the effect of radiation pressure on the ISN H measurements for the cases with and without momentum transfer and compared them with our previous findings. We found an effective radiation parameter ($\mu_{\scriptsize\textrm{eff}}$, which represents for force associated with radiation pressure relative to gravity) for the years 2009-2018 based on the longitudinal shift of the ISN H signal. The two cases with and without momentum transfer reproduce the longitudinal shift in accordance with variations in solar activity, in agreement with our previous results, and they result in similar values for the $\mu_{\scriptsize\textrm{eff}}$, which is $\sim21-22 \%$ larger than the value found based on Ly$\alpha$ observations.
\end{abstract}


\section{Introduction} \label{sec:intro}

IBEX-Lo is designed to detect interstellar neutral (ISN) atoms from the vantage point of elliptical orbits around the Earth, i.e., at 1 au \citep{McComas2009, Fuselier2009}. The ISN observation season occurs each year from December through April when the Earth and IBEX move into the approaching ISN flow, increasing the incident atoms’ relative velocity, and thus energy in the spacecraft reference frame. This mission is designed so that the instrument’s field of view remains perpendicular to near Sun-pointing (with maximum $ \sim 4 \degr -7 \degr$ off for each orbit) spin axis. As a consequence, ISN atoms are expected to be detected close to their trajectory perihelion so that their radial velocity is close to zero as they enter the instrument’s narrow field of view. The angular distribution of observed ISN fluxes is essential for analyzing ISN flow parameters.

Our very local interstellar medium (VLISM) is believed to be an interaction region between the local interstellar cloud (LIC) and galactic (G) cloud \citep{Swaczyna2022}. Therefore, it is important to investigate the VLISM through its messengers (the ISN atoms and ENAs, energetic neutral atoms). Investigating parameters associated with the ISN flow has been important for understanding the VLISM \citep{Mobius2004}. A Maxwellian may adequately describe the ISN distribution in the VLISM far away from the Sun, where its ionized and neutral parts are highly coupled through charge exchange and elastic collisions. However, encountering the heliosphere heats, decelerates, and diverts the plasma part, forming a bow shock \citep[or bow wave, see][]{McComas2012, Zank2013, Fraternale2023}. The neutral part, on the other hand, enters the outer heliosheath without significant changes. 

Charge exchange and elastic collisions between ions and neutral atoms in the outer heliosheath are not sufficient to maintain thermalization, leading to decoupling the plasma and neutral part. The charge exchange collisions lead to depletion of the pristine (primary) population and creation of a secondary population with different thermodynamic properties. Previous models excluded the effect of momentum transfer (hereafter MT) between ions and neutral atoms, assuming it to be negligible. However, for ISN He, this MT significantly alters the distribution function of both primary and secondary populations, particularly leading to a noticeable deviation from the pristine Maxwellian distribution for the primary ISN He \citep{Swaczyna2023}. 

Inside the heliosphere, modification of the ISN atoms' distribution function by the gravitational force from the Sun and ionization processes continues and increases closer to the Sun. Radiation pressure exerted on the ISN H atoms due to resonant absorption and reemission of Ly$\alpha$ further decelerates them. Therefore, while ISN He, O, and Ne atoms have the same bulk speed and their mean energies at 1 au scale with their mass, the only exception is ISN H. Various He observations obtained the ISN inflow speed $\sim 25$ km s\textsuperscript{–1} at the edge of the heliosphere \citep{Witte2004, Bzowski2014, Wood2015, Vallerga2004, Bzowski2015, McComas2015, Schwadron2015}. The other parameters that describe the ISN flow distribution function at the edge of the heliosphere include the temperature, density, and direction \citep{Lallement1992, Linsky1993, Gloeckler2004, Mobius2004}. 

IBEX provided the first direct measurements of ISN H \citep{Mobius2009} and the first measurements of ISN H and He throughout a full solar cycle \citep{Galli2019, Swaczyna2022, Schwadron2022}. The analysis of ISN He observations from IBEX has provided the latest estimates for the unperturbed ISN parameters as reported by \citet{Swaczyna2022, Schwadron2022}. The analysis of ISN H data over several years \citep{Saul2012, Saul2013} and a full solar cycle \citep{Galli2019, Rahmanifard2019} showed their variations with solar activity. Furthermore, investigating the ISN signal over different phases of solar activity demonstrated the effect of radiation pressure on the ISN H signal \citep{Schwadron2013, Katushkina2015, Kowalska2018b}, namely a longitudinal shift of the ISN H signal with increasing solar activity \citep{Zirnstein2013, Rahmanifard2019, Katushkina2021}.

ISN H and He are both detected by IBEX-Lo through registering H$^-$ ions either converted from incoming ISN H atoms or sputtered from the conversion surface \citep{Wurz1997, Wurz2008, Fuselier2009, Swaczyna2023_HeResponse}. Therefore, retrieving ISN H count rates from H$^-$ ions presents a challenge and requires further data analysis. This analysis is mostly focused on the effect of radiation pressure, which shifts the ISN H signal in longitude and separates the two signal peaks \citep{Schwadron2013}. While this separation is present throughout different phases of a solar cycle, it increases with solar activity. \citet{Galli2019} explored several retrieval methods by analyzing H$^-$ and O$^-$ count rates during different phases of the observation season when we expect to observe secondary He, primary He, and H signals. They confirmed the basic finding of \citet{Saul2012, Saul2013} and \citet{Schwadron2013} i.e. that the observation of higher ISN H count rates in energy step (E-Step) 1 (centered at 15 eV) than in E-Step 2 (29 eV) does not depend on the retrieval method.  

The larger-than-unity ratio between E-Step 1 and 2 has led to a discrepancy between IBEX data and existing models \citep{Katushkina2015, Katushkina2021, Galli2019, Rahmanifard2019}. While increasing the radiation parameter \citep{Kowalska2018a, Kowalska2018b, Kowalska2022} compared to solar Ly$\alpha$ observations (possibly associated with an absolute calibration issue with solar Ly$\alpha$ irradiance observations) may partially address this discrepancy, it is still unresolved. To sidestep the discrepancy, in this study and in \citet{Rahmanifard2019}, we have focused on the longitudinal shift of the ISN H signal and its variation with solar activity. In Section \ref{sec:heliosheath}, we investigate collisions of ISN H atoms in the outer heliosheath and how they alter the distribution of H atoms at a fiducial boundary of 100 au from the Sun (i.e., near the solar wind termination shock). We provide a summary of our global model and compare the two cases with and without considering MT. In Section \ref{sec:heliosphere}, we introduce our simulation tool inside 100 au (analytical full integration model, aFINM). We briefly describe how it has been modified and is paired with the global model for this study. We further discuss the results of our ISN H analysis through the years 2009-2018 and compare our results for three cases where we used the 2019 version of the aFINM (hereafter aFINM2019), and the modified version with and without MT in Section \ref{sec:heliosphere}. We provide a brief discussion and concluding remarks in Section \ref{sec:summary}.

\section{Collisions of ISN Hydrogen Atoms in the Outer Heliosheath} \label{sec:heliosheath}

In this study, we closely follow the methodology developed to model collisions of ISN helium atoms by \citet{Swaczyna2021, Swaczyna2023}. We simulate the transport of ISN atoms through the outer heliosheath using a Monte Carlo simulation of test particle trajectories, employing flow velocities and temperatures of plasma and neutral components obtained from the global heliosphere model \citep{Zirnstein2016}. The global heliosphere model itself does not account for the MT related to the angular scattering of atoms and ions, and the results of the test particle calculations are not reflected in the flows and temperatures. As such, the calculations are not self-consistent. However, \citet{Heerikhuisen2009} concluded that MT related to charge exchange collisions does not significantly influence the global heliosphere structure or flows of neutral and charged components. Therefore, the higher-order effects should not invalidate the conclusions of this study. This section briefly describes our methodology, but a more detailed explanation can be found in \citet{Swaczyna2023}. 

The calculations start at the outer boundary, 500 au from the Sun, where we randomly select the initial state vectors of ISN hydrogen atoms from the Maxwell distribution. Then, we calculate the probabilities of charge exchange and elastic collisions with protons which depend on the cross-sections for collisions and the protons distribution in the global model along the trajectory of each atom. These probabilities are used to randomly select whether a collision occurred within each step. The step size is defined so that the combined collision probability does not exceed $5\%$ and the trajectory length within the step is less than 3 au. If the collision occurs, the atom’s velocity is modified using the differential cross section \citep{Schultz2016} as the probability distribution function. The atoms are tracked until their distance from the Sun is less than 100 au (inner boundary) or until they cross the plane that contains the Sun and is perpendicular to the flow velocity vector far from the Sun (the downwind hemisphere of the heliosphere). During their journey between the outer and inner boundary, we neglect radiation pressure because, according to \citet{Kowalska2022}, the part of the solar Ly$\alpha$ line responsible for radiation pressure on ISN hydrogen atoms is almost fully absorbed beyond 100 au. 

In contrast to helium, charge exchange and elastic collisions between hydrogen atoms and protons are the two dominant types of collisions of hydrogen atoms in the outer heliosheath. Collisions with He$^+$ ions are an order of magnitude lower due to $\sim 2-3$ times smaller cross section and $\sim 6$ times lower density of He$^+$ ions than protons \citep{Bzowski2019}. Furthermore, we do not account for elastic collisions between hydrogen atoms because they would require a global heliosphere model that accounts for these collisions self-consistently. Nevertheless, these collisions do not transfer momentum between populations but lead to the internal thermalization of ISN hydrogen atoms. The relevant cross section for such a study can be found in \citep{Ovchinnikov2017}. While a detailed study of this problem is beyond the scope of this paper, the frequency of these collisions is insufficient to fully thermalize the ISN hydrogen population.

Based on the above conditions, we find the properties of atoms at the inner boundary as a function of their entrance position at 100 au from the Sun using the procedure described in \citet{Swaczyna2023}. First, we determine the density and bulk flow velocity of the primary and secondary ISN hydrogen populations as a function of the position at the inner boundary. In the next step, we characterize the 3D distribution functions using components parallel and perpendicular to the bulk flow velocity in the solar frame separately for each population. The obtained distribution functions are discussed in Appendix A. Table \ref{tab:pop_prop} shows a summary of the properties of these populations. The table shows the results obtained with and without including the MT (angular scattering). In both cases, the charge exchange collisions are accounted for, but for the calculations without MT, the speed of the newly neutralized atoms is adopted from the velocity of the parent proton. 

\begin{table}
\caption{\label{tab:pop_prop} Properties of ISN hydrogen populations at 100 au}
\centering
\begin{ruledtabular}
\begin{tabular}{ccccccc}
{Population} & {MT} & {$n/n_\infty$} & {$v$ (km s\textsuperscript{–1})} & {$\alpha (\degr)$} & {$T_\parallel (K)$}  & {$T_\perp (K)$} \\

\hline
Model VLISM & - & 1 & 25.4 & 0 & 7500 & 7500 \\
 \multirow{ 2}{*}{Primary}& With & 0.224 & 26.5 & +1.3 & 7540 & 8440 \\
 & Without & 0.229 & 27.8 & +0.5 & 6160 & 7540 \\
 \multirow{ 2}{*}{Secondary}& With & 0.521 & 16.8 & +8.8 & 10130 &  11020 \\
 & Without & 0.508 & 16.6 & +7.8 & 9860 & 11050 \\

\end{tabular}
\end{ruledtabular}
\end{table}

\begin{figure*}
\gridline{\fig{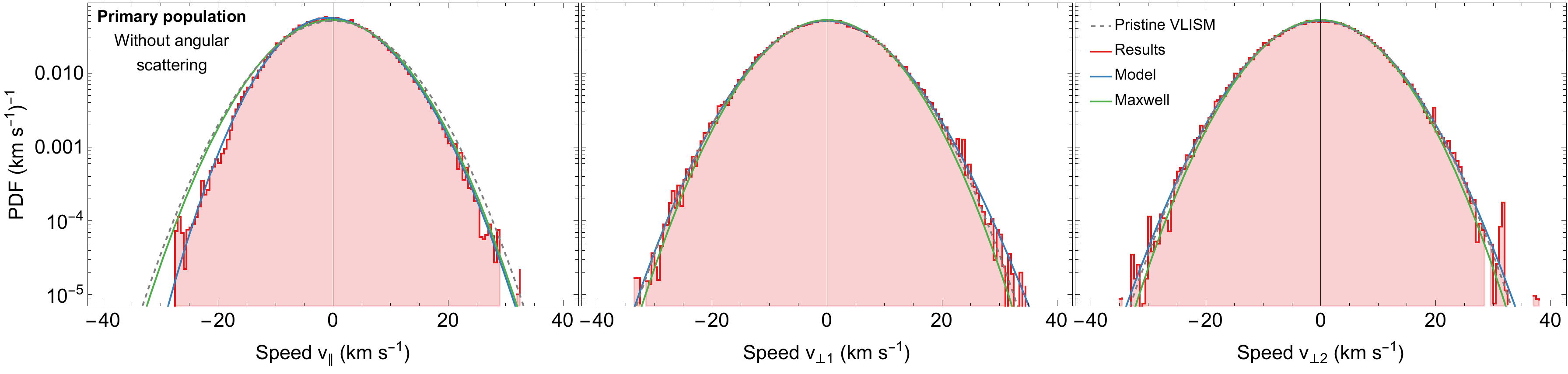}{1.0\textwidth}{}
          }
\gridline{\fig{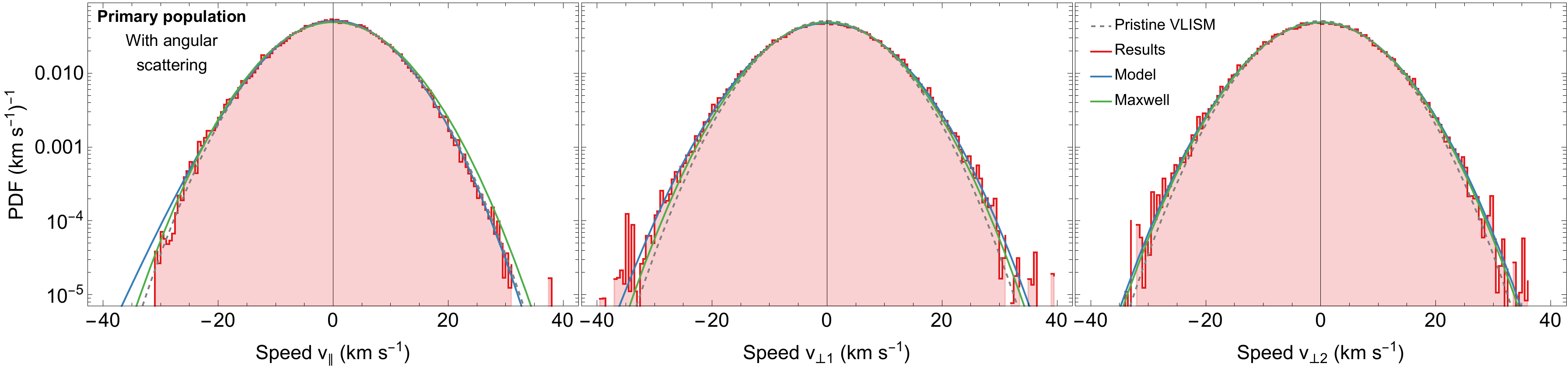}{1.0\textwidth}{}
          }
\caption{Distribution functions of the primary population of ISN H. Panels from left to right show the parallel and two perpendicular components. The bulk speed is subtracted from the parallel component, and positive parallel speeds correspond to atoms moving faster than the bulk flow toward the Sun. The top and bottom rows present the results obtained without and with accounting for the angular scattering in collisions (MT). The red histogram presents the results of the Monte Carlo simulations, and the blue and green lines show the analytic model derived in Appendix A, and the Maxwell approximation, respectively. The original distribution in the VLISM is shown with a dashed gray line. \label{fig:pripop}}
\end{figure*}

\begin{figure*}
\gridline{\fig{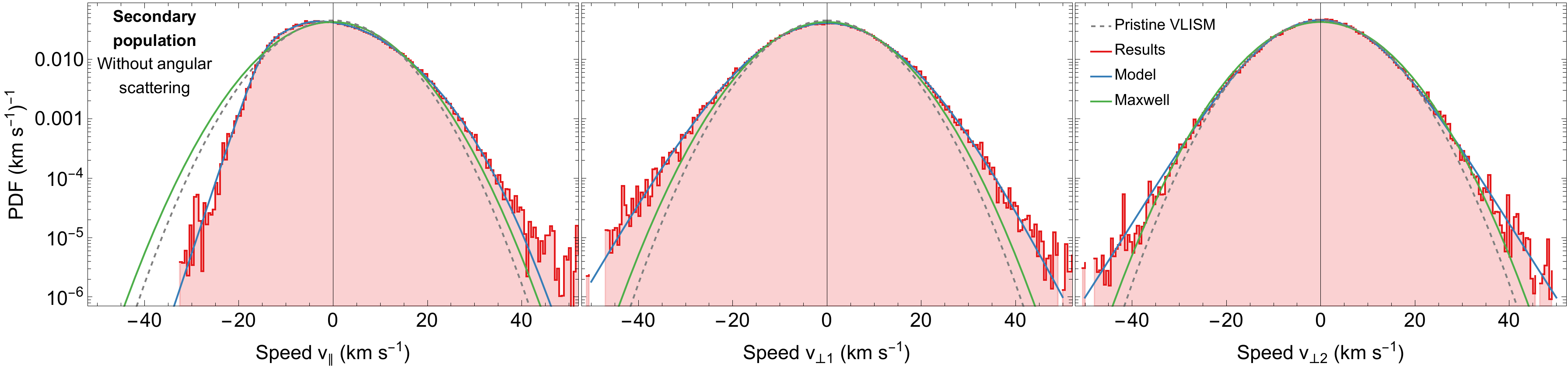}{1.0\textwidth}{}
          }
\gridline{\fig{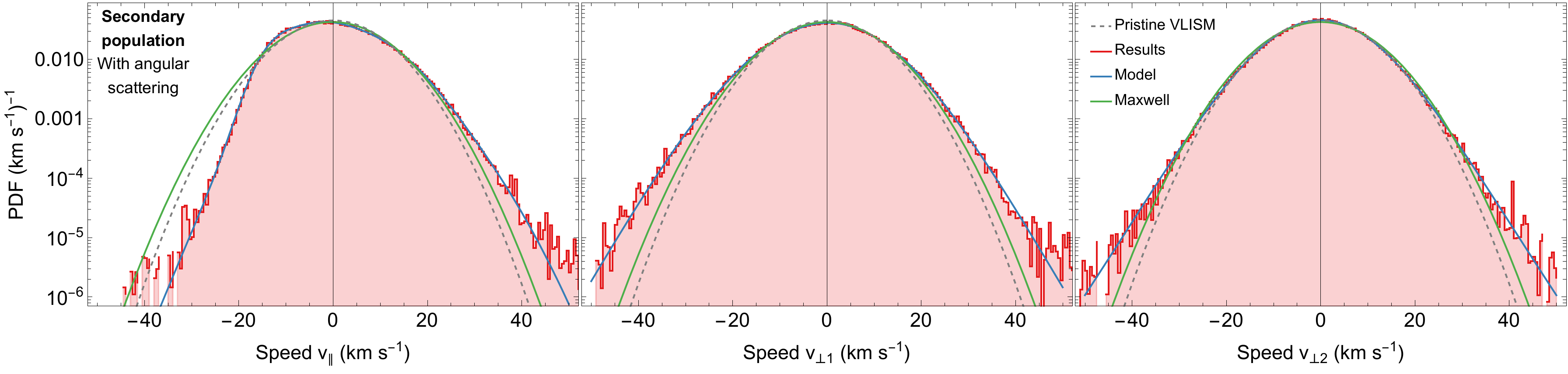}{1.0\textwidth}{}
          }
\caption{As in Figure \ref{fig:pripop}, but for the secondary population.  \label{fig:secpop}}
\end{figure*}

The table shows the relative density of each population compared to the density of ISN hydrogen far from the Sun ($n/n_\infty$), the bulk speed of atoms along the inflow direction ($v$), the deflection angle from the flow direction far from the Sun ($\alpha$), and temperature of the distribution in the direction parallel ($T_\parallel$) and perpendicular ($T_\perp$) to the flow direction. The temperatures provided here are calculated from the mean thermal speed. Figures \ref{fig:pripop} and \ref{fig:secpop} present the distribution functions of the parallel and perpendicular velocity components in the frame oriented along the bulk flow at 100 au. The figure compares the histograms obtained from the Monte Carlo simulations, the approximation using the Maxwell distributions with temperatures provided in Table \ref{tab:pop_prop}, and the analytic distribution functions derived in Appendix A.

Unlike ISN helium \citep{Swaczyna2023}, the primary population of ISN H in each direction can be relatively well represented by Maxwell distribution with temperatures presented in Table \ref{tab:pop_prop} (see Figure \ref{fig:pripop}). However, the secondary population requires a superposition of four Maxwellian populations to capture the complexity of the parallel component and asymmetric kappa distributions \citep{Swaczyna2021} to model the perpendicular components (see Appendix A). 

The modeling results without the MT show that the primary population is apparently "accelerated" before it reaches 100 au. While we include gravity, it is responsible for a speed increase by only $\sim 0.4$ km s\textsuperscript{–1} compared to the pristine VLISM. The remaining difference is due to the charge exchange process more effectively removing primary atoms moving slower relative to the Sun because the cross section increases for lower collision speeds \citep{Heerikhuisen2016, Bzowski2020}. The same effect causes a reduction of the parallel temperature of this population. The preferential removal of atoms by charge exchange collisions also deflects the bulk velocity by $\sim +0.5\degr$ away from the flow direction within the B-V plane defined by the interstellar flow velocity and magnetic field direction. The positive value of this angle means that the apparent inflow direction is closer to the interstellar magnetic field direction. The inclusion of the MT slows down the population by $\sim 1.3$ km s\textsuperscript{–1}, increases the deflection by $\sim 0.8\degr$, and increases temperature by $\sim 1100-1300$ K. 


In both cases, the secondary population obtained in our analysis is slower, more deflected within the B-V plane, and warmer than the primary population. However, the MT effects are much weaker than the estimate for the ISN helium populations; the speed increases by just $\sim 0.2$ km s\textsuperscript{–1}, the deflection angle increases by $\sim 1\degr$, and the temperature remains almost the same. These changes differ from the conclusions of \citet{Swaczyna2019}, who showed that the secondary population obtained from the charge exchange of neutral hydrogen atom and proton populations moving with respect to each other with a relative bulk speed of $\sim 20$ km s\textsuperscript{–1} result in a secondary population moving faster relative to the Sun by $\sim 2$ km s\textsuperscript{–1} if the MT is accounted for. They also predicted that the population would be substantially warmer. However, the analysis by \citet{Swaczyna2019} considered idealized charge exchange collision of two Maxwell populations, neglecting any subsequent collisions of secondary atoms in the outer heliosheath. Our current analysis shows that the atoms that undergo one charge exchange collision are subsequently more likely to undergo additional collisions, including another charge exchange, than the primary atoms. Therefore, subsequent collisions reduce the MT effect because the typical relative collision speed between a secondary atom and proton is much lower. 

The primary population is composed of atoms that are not subject to charge exchange collision between the outer and inner boundaries. On the other hand, the secondary population is formed by those created through charge exchange in this region. The distance of the outer boundary (500 au) is adopted so that the plasma and neutral flows beyond this distance are close to their pristine conditions. Nevertheless, the secondary atoms from charge exchange collisions near the outer boundary have properties close to the primary atoms and could also be classified as such. As we discussed in \citet{Swaczyna2023}, this would not impact the combined population as long as we consider the sum of these two populations, but any comparisons of the primary and secondary population properties need to account for this somewhat arbitrary separation of the two populations. The significance of the distance of the inner boundary is less important as long as it is located within the heliopause, so we can assume that the accumulation of the secondary population ceased. The charge exchange collisions between the heliopause and inner boundary may result in energetic neutral atoms with energies substantially exceeding those of the primary and secondary neutrals. Therefore, we remove them from further analysis.

\section{Modeling ISN H inside 100 AU Boundary} \label{sec:heliosphere}

The analytical full integration model (aFINM) is an attempt to integrate the analytic relations based on the classical hot model adapted by \citet{Lee2012} to the IBEX observation strategies. These observation strategies are introduced to our simulations through three numerical integrations over the instrument response function and spin sector to calculate ISN count rates observed by IBEX-Lo. This model has been previously used to characterize the ISN H \citep{Schwadron2013, Rahmanifard2019}, ISN He \citep{Schwadron2015} and ISN O \citep{Schwadron2016} flow. In these studies, as in \citet{Lee2012, Lee2015} a drifting Maxwellian distribution is used for the primary population of ISN atoms at infinity:

\begin{equation}
\label{eq:1}
f_{i} = n_{i} (2 \pi k_B T_{i} / m)^{-3/2} \times \exp(- \mid \vec{V}_\infty - \vec{V}_{i} \mid^2 (2 k_B T_{i} / m)^{-1}) \
\end{equation}  

where $ \vec{V}_\infty$ is the velocity for an ISN atom with a given energy at a specific observation geometry traced back to the termination shock (which we assumed to be located at infinity in aFINM2019). $n_i$, $T_i$, and $\vec{V}_i$ are the density, temperature, and bulk flow velocity for the primary ISN atoms. In \citet{Rahmanifard2019}, we further considered a secondary population for ISN H atoms with a drifting Maxwellian as described by Eq (1) so that the distribution function includes both primary and secondary populations:

\begin{equation}
\label{eq:2}
f = \sum_{i=prim,sec} f_{i} 
\end{equation}

where we adopted parameters associated with primary and secondary populations from previous studies \citep[][etc.]{Bzowski2008, Kubiak2016}.

In this study, however, we consider the transport of ISN atoms through the outer heliosheath based on the global heliosphere model \citep{Zirnstein2016}. As described in Section \ref{sec:heliosheath}, the transport of ISN atoms in the heliosheath is considered between 500 au and an inner boundary at 100 au, where aFINM takes over the resulting distribution function. This 3D distribution function is a combination of ISN primary and secondary populations. Therefore, we have modified the aFINM model to accept distribution functions that are a function of the position relative to the Sun, as well as the velocity. To make this change, we used a different set of equations for the trajectory of ISN atoms, as shown in Appendix A of \citet{Swaczyna2023}. We further made two modifications to account for the radiation pressure: first, the standard gravitational parameter of the Sun was replaced with an effective gravitational parameter $\alpha = (1 - \mu) GM_\odot$, where $GM_\odot = 1.3271244 \times 10^{22}$; additionally, the eccentricity sign adopted in their equations was changed for a repulsive force, i.e., when $\mu > 1$. 

With the above changes, we calculate the position and velocity vectors at the 100 au inner boundary for a given energy at a specific IBEX observation geometry. Based on these values we find the distribution function value at the inner boundary which is further multiplied by a survival probability that is calculated based on photoionization and charge exchange probabilities to obtain the distribution function at 1 au \citep[see][for further details]{Rahmanifard2019, Sokol2020, Sokol2013, Sokol2014}. We then calculate the flux of ISN H atoms and integrate over the response function of the instrument including energy and collimator response functions to find the observed count rates at a specific time and specific spin angle. We further integrate over a full spin bin and good times list of an orbit to provide the observed count rates associated with that orbit and spin bin. 

Our main purpose here is to find out how considering the MT caused by charge exchange and elastic collisions in the heliosheath affects the results we obtained from Rahmanifard (2019). In \citet{Rahmanifard2019}, we concentrated on the longitudinal shift in the peak of the ISN H signal and found an effective radiation parameter that matched this shift for each year. This constraint on the analysis was mainly due to a previously reported discrepancy between IBEX data and existing models. This discrepancy was first reported by \citet{Katushkina2015}, where they showed that, contrary to IBEX data, their model obtains higher count rates for ISN H in E-Step 2 than E-Step 1, in the expected radiation pressure and ionization rate range. The same discrepancy has been reported in comparison with other models including aFINM2019 \citep{Rahmanifard2019} and WTPM \citep{Galli2019}, which limits the analysis of ISN H intensity. Therefore, in \citet{Rahmanifard2019}, we developed a methodology to use the longitudinal shift in the ISN H signal peak to explore the effect of radiation pressure on ISN H signal. The longitude, in this paper, represents the ecliptic longitude of IBEX, from the fall equinox, at the time of observation. 

As detailed in \citet{Rahmanifard2019}, we found an effective value for the radiation parameter ($\mu_{\scriptsize\textrm{eff}}$) with the best agreement between the observed and model-predicted $\lambda{\scriptsize_\textrm{peak}}$ (the longitude at which the ISN H signal peaks). To find the observed $\lambda{\scriptsize_\textrm{peak}}$, we first ran a $\chi^2$ minimization analysis to fit a Gaussian to the observed ISN H signal each year. Our focus was on IBEX-Lo spin sectors 14 and 15, which correspond to $6\degr$ bins with their centers located at $84\degr$ and $90\degr$ from the north ecliptic pole (NEP). We used these two spin bins because they are closest to the latitudinal peak of the signal and provide higher statistics. Furthermore, we used the signal observed in E-Step 1 because it is the only energy channel where we can observe ISN H throughout almost the entire solar cycle. While ISN H signal is present in E-Step 2 for the first years of IBEX observations, count rates in E-Step 2 are statistically zero for all the remaining years. This decrease is partly due to moving toward a solar maximum, and partly due to lowering the post-acceleration (PAC) voltage from 16 kV to 7 kV in the summer of 2012.

While the focus of this study is on the longitudinal shift of the signal peak, strong variations of the intensity of the signal with solar activity affects the analysis. Moving toward a solar maximum caused the relatively noticeable signal associated with the years 2009-2011 to decrease to values close to the background level in 2012. The signal became undetectable for the years 2013 to 2015 mostly due to extensive ionization and radiation pressure during solar maximum. Also, the reduction in the PAC voltage lowered the statistics by half. We further missed the signal for the year 2016 due to a different stepping scheme. Therefore, the focus of \citet{Rahmanifard2019} and this study is on the years 2009-2012 and 2017-2018. We further added 2013 and 2014, without running a $\chi^2$ analysis since there are too few data point to run a $\chi^2$ analysis. Instead, we simply used the highest data point as the peak of the signal for these two years. 

We then performed a set of $\chi^2$ analyses to fit a Gaussian to our model-predicted count rates to find values for $\lambda{\scriptsize_\textrm{peak}}$ from our model for different values of the radiation pressure parameter. Running this analysis iteratively, we found $\mu$ values which matched $\lambda{\scriptsize_\textrm{peak}}$ from the observations. The effective $\mu$ values ($\mu_{\scriptsize\textrm{eff}}$) we obtained in this way are empirical. 

\begin{figure}[ht!]
\plotone{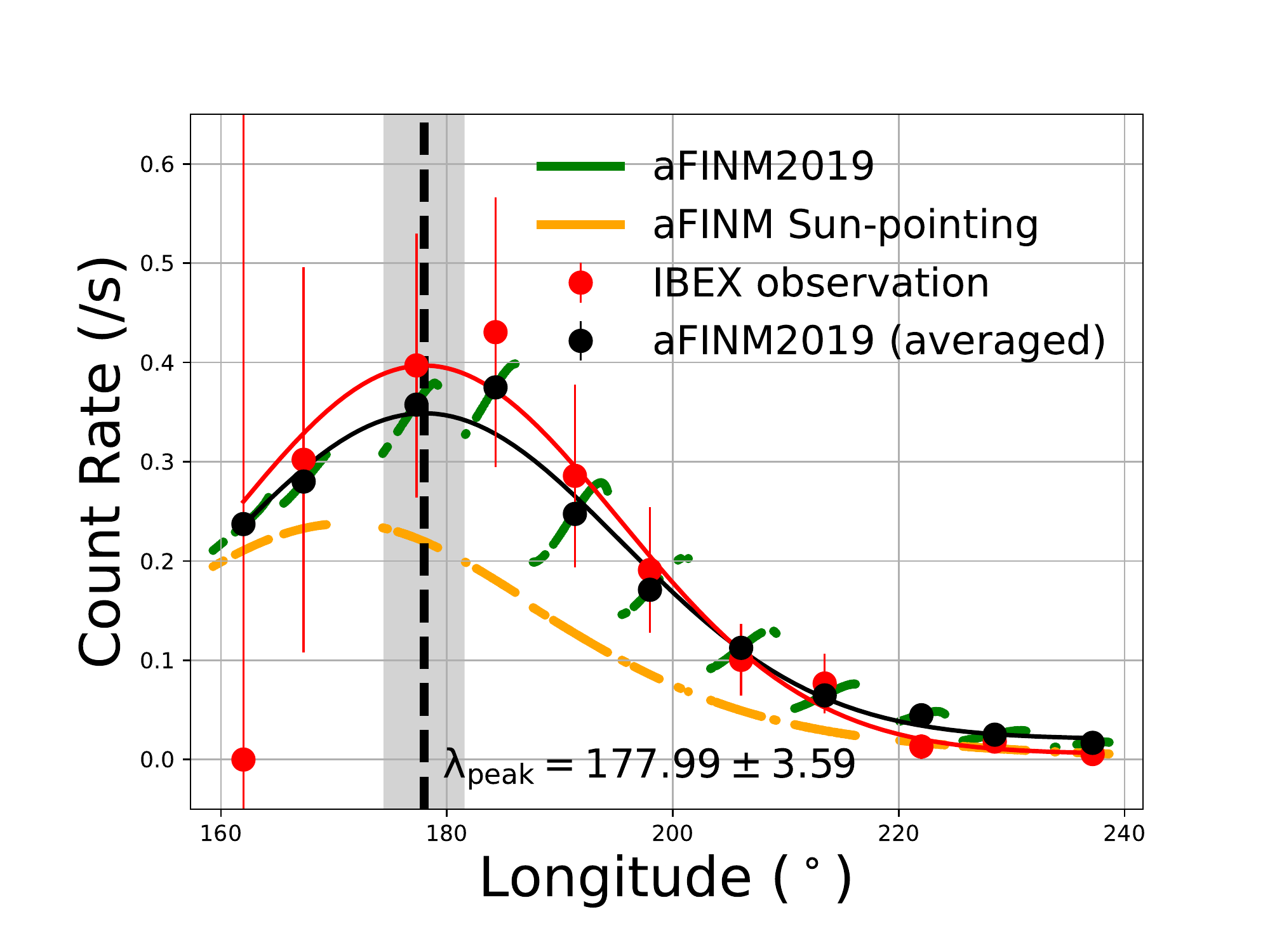}
\caption{Model predictions based on aFINM2019 compared to IBEX-Lo observations of ISN H for the year 2011 (E-Step 1, spin bin 14) adopted from Rahmanifard 2019. The green and orange line show model predictions for actual IBEX observations and Sun-pointing Earth location and reference frame. The black line is the Gaussian fit to model predictions averaged over the times when IBEX data is available (black data points) and the red line is the Gaussian fit to IBEX data (red data points). The dashed black line shows the peak longitude of the ISN H signal and gray area shows the uncertainty region. \label{fig:model}}
\end{figure}

\begin{figure}[ht!]
\plotone{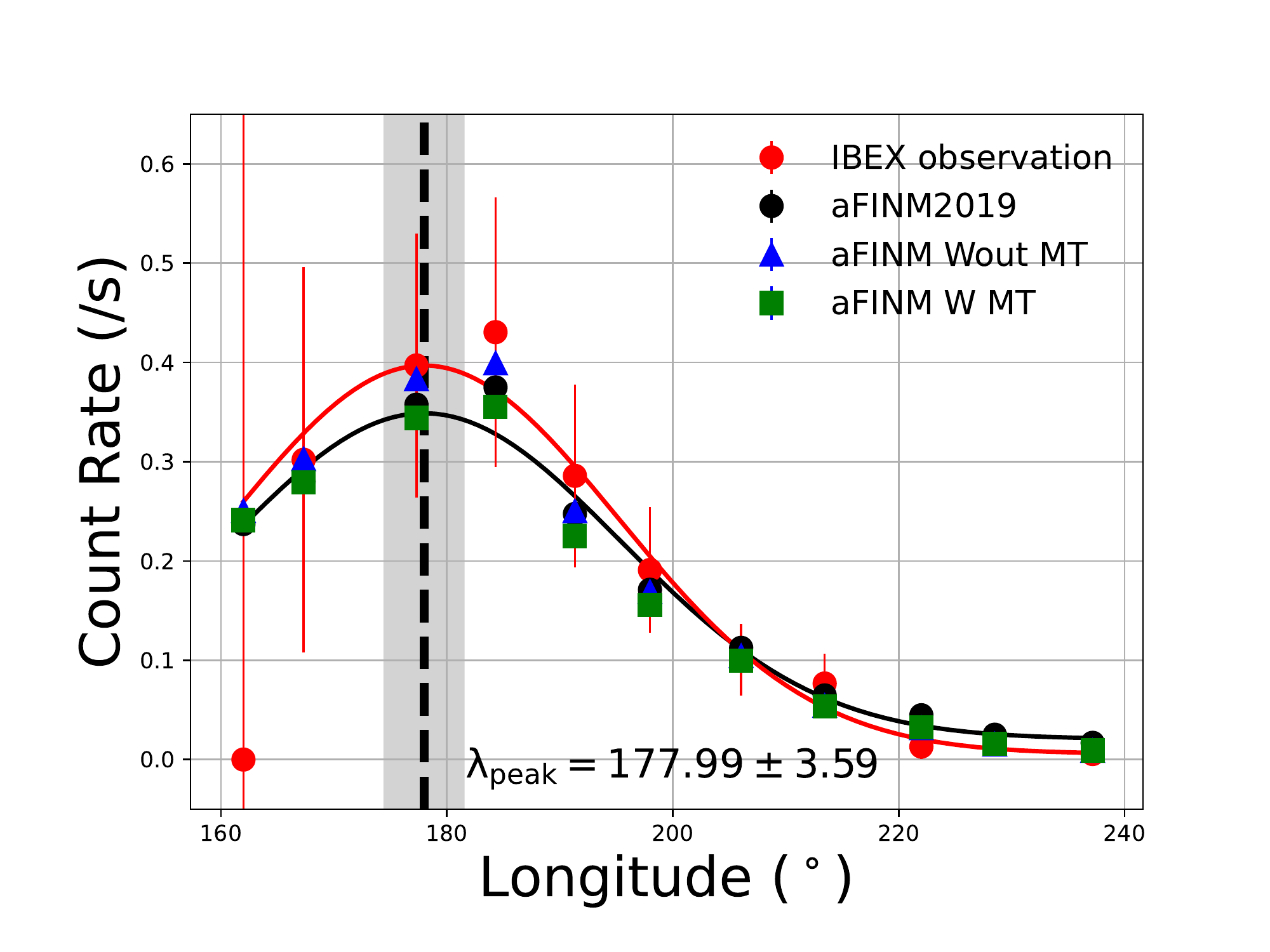}
\caption{Model predictions based on aFINM2019 (black circles) and modified aFINM for the two cases with (green squares) and without (blue triangles) MT compared to IBEX-Lo observations (red circles) of ISN H for the year 2011 (E-Step 1, spin bin 14). The black line is the Gaussian fit to aFINM2019 and the red line is the Gaussian fit to IBEX data. The dashed black line shows the peak longitude of the ISN H signal and gray area shows the uncertainty region.  \label{fig:mupdate}}
\end{figure}

Using a Gaussian fit to find $\lambda{\scriptsize_\textrm{peak}}$ for the observed and predicted ISN H signal is a reasonable method. Figure \ref{fig:model}, which we have adopted from Figure 4 in \citet{Rahmanifard2019} shows aFINM2019 predictions for IBEX observations in spin sector 14 in 2011 ISN H season. The green line shows aFINM2019 predictions for IBEX observations throughout each orbit using exact spacecraft's position, velocity, and spin axis. The orange line, on the other hand, represents the Earth-located aFINM2019 predictions in the Earth reference frame with the spin axis of the instrument Sun-pointing at all times. The black circles show the predicted count rates (the green line) averaged over the observation times and the red circles with error bars show the the observed count rates. The black (red) line presents a Gaussian fit to the black (red) circles. Comparing the green and orange lines demonstrates the effect of the spacecraft motion and the field-of-view drifts relative to the ideal Sun-pointing spin axis creating the sawtooth patterns. In addition to these effects, differently distributed observation times for each orbit leads to a deviation from the Gaussian fit, especially for orbit 118 (the forth orbit from left) in Figure \ref{fig:model}. Despite such deviations, using a Gaussian fit in our analysis is reasonable since we include these effects in our model and use the same techniques to find the peak for the observed and predicted signal. It is important to note that the absolute location of $\lambda{\scriptsize_\textrm{peak}}$ does not play a role in our analysis. Therefore, as can be seen in Figure \ref{fig:model} and Figure \ref{fig:mupdate}, we use the peak of the Gaussian fit (dashed black line with gray uncertainty region) rather than the maximum measured count rate.

Figure \ref{fig:mupdate} shows our results for the year 2011 in sector 14 as an example to compare our aFINM2019 results with the current version of the code in which we consider charge exchange and elastic collisions in the heliosheath. All aFINM simulation results presented in Figure \ref{fig:mupdate} are based on $\mu_{\scriptsize\textrm{eff}}$ as obtained in our previous study for the year 2011 ($\mu_{\scriptsize\textrm{eff}}=1.10424$). We present the observed count rates for each orbit as red circles with error bars and the predicted aFINM2019 rates as black circles. We further include our modified version of aFINM, with (green squares) and without (blue triangles) MT from charge exchange and elastic collisions. 

The main difference between aFINM2019 and aFINM without MT in this figure is the selective charge exchange in the outer heliosheath (as detailed in Section \ref{sec:heliosheath} and Appendix A). Additionally, using the inner boundary at 100 au rather than using pristine parameters at infinity for the primary population leads to slightly different trajectories inside the heliosphere as described earlier. As discussed in Section \ref{sec:heliosheath}, the selective charge exchange leads to a faster and cooler primary population. The results, as can be seen in Figure \ref{fig:mupdate}, do not show a significant shift in the peak location between aFINM2019 and the modified version of aFINM without MT, except for a small relative increase close to the peak compared to the flanks of the signal. This difference may arise from the fact that in the new version the speed difference between the primary population (dominant in the peak) and the secondaries (dominant in the flanks) is even more pronounced than in aFINM2019. 

The green squares in Figure \ref{fig:mupdate} account for the MT both in the charge exchange and elastic collisions. The selective charge exchange of the primary H population in the outer heliosheath is present for this case (with MT) as well. However, as explained in Section \ref{sec:heliosheath}, accounting for the MT in the charge exchange and elastic collisions slows down the primary ISN H significantly, thus the predicted signal for this case shows lower count rates close to the peak. 


\begin{figure}[ht!]
\plotone{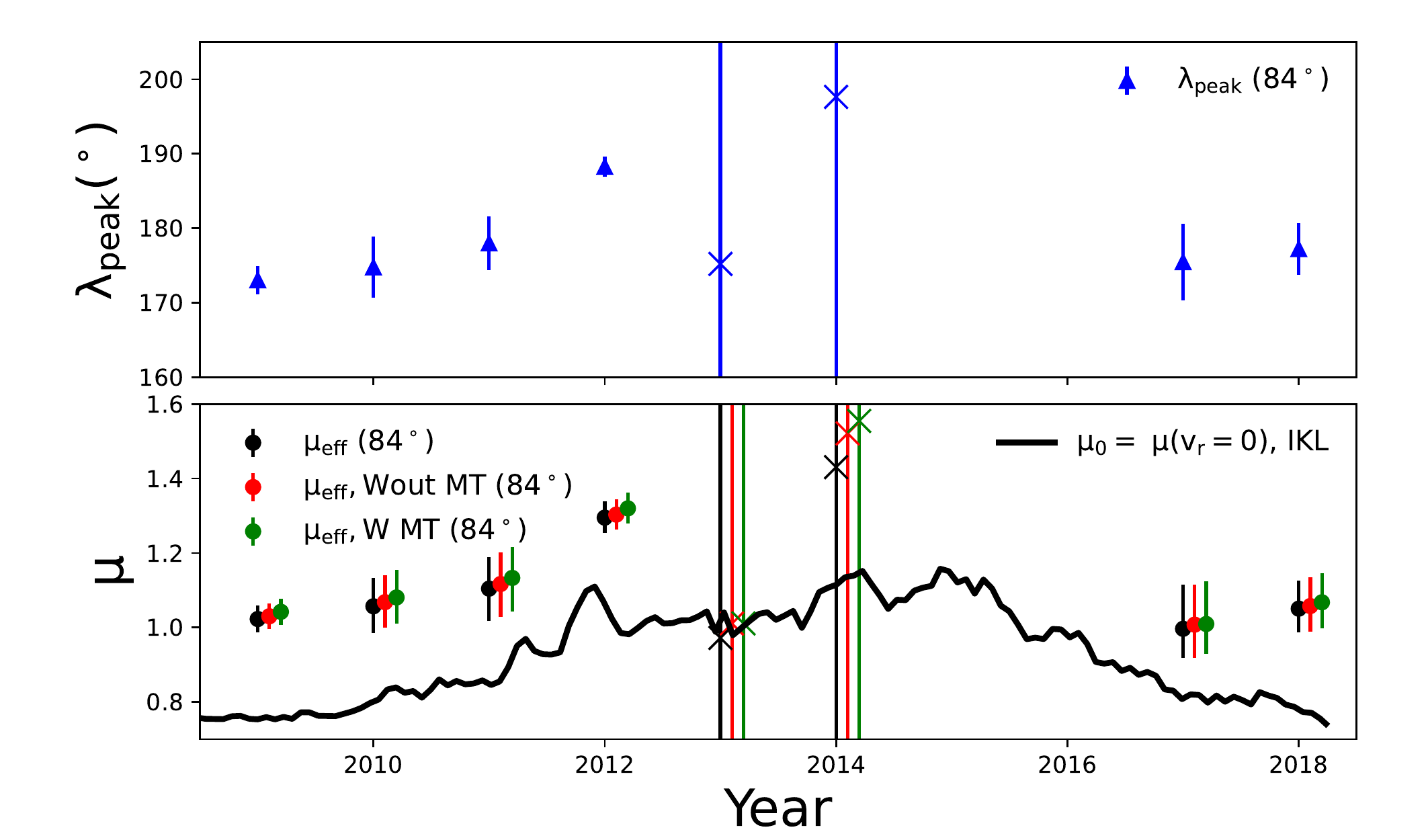}
\caption{ISN H flow peak longitude based on IBEX observations at spin bin 14 ($84\degr$ from the NEP). Blue triangles with error bars are found by fitting a Gaussian function to ISN H signal for the years 2009-2012 and 2017-2018. Blue x marks used for the years 2013 and 2014 are obtained based on the highest observed data points. Bottom panel: $\mu_{\scriptsize\textrm{eff}}$ obtained based on our analysis for aFINM2019 (black), modified aFINM without MT (red), and with MT (green) circles with error bars for spin bin 14 data. The black line represents $\mu_{\scriptsize\textrm{0}}$ based on \citet[][IKL]{Kowalska2018a}. \label{fig:mu14}}
\end{figure}

\begin{figure}[ht!]
\plotone{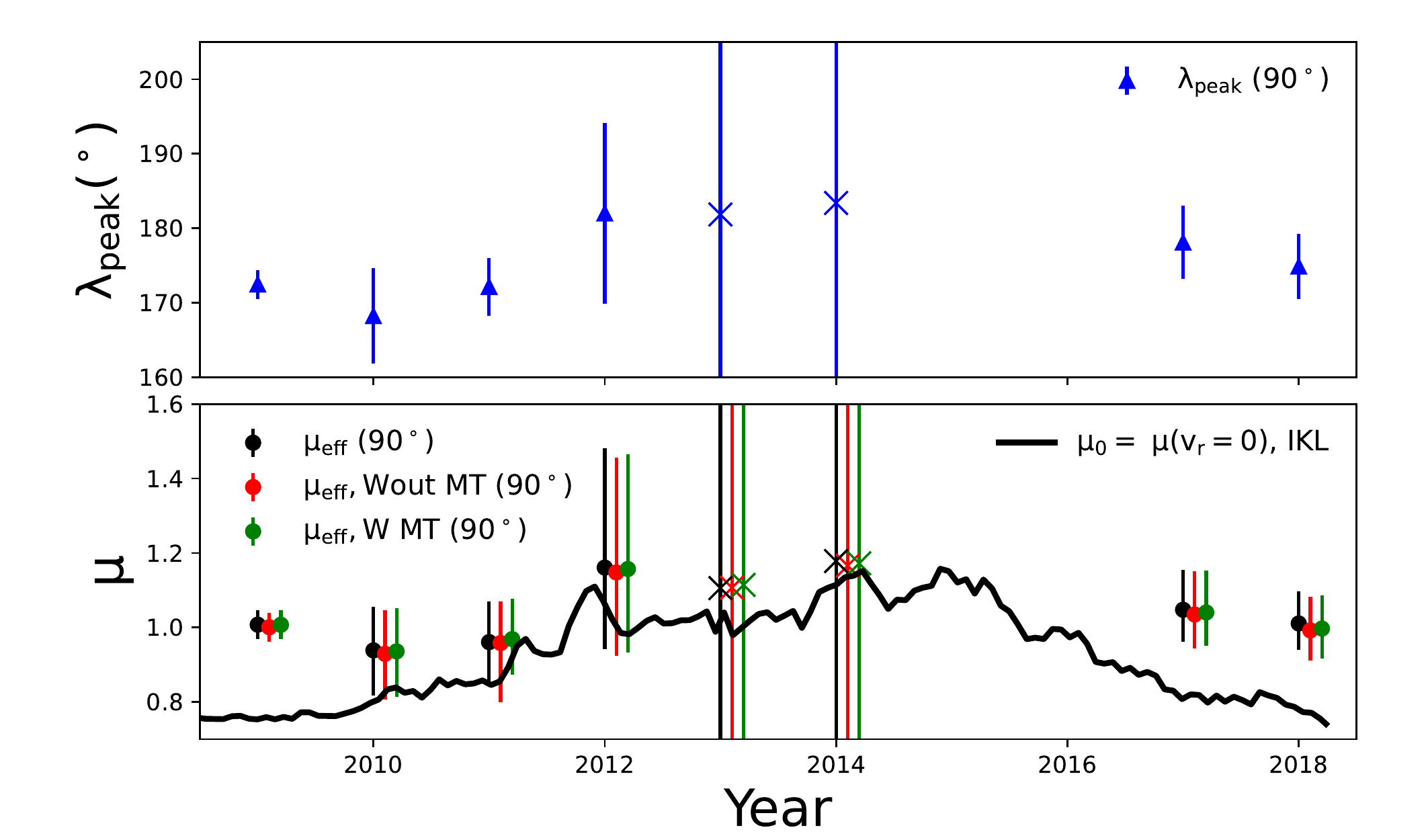}
\caption{As in Figure \ref{fig:mu14} but for spin bin 15 ($90\degr$ from the NEP).\label{fig:mu15}}
\end{figure}

We performed this analysis for the years 2009-2018 (spin bins 14 and 15), excluding 2015-2016, with the new version of the code, for the two cases with and without MT. We present the results in Figures \ref{fig:mu14} (spin bin 14) and \ref{fig:mu15} (spin bin 15). We see the same pattern as in aFINM2019, with slight variations in the $\mu$ values. Despite these small variations, the same trend still persists. Black, red, and green circles with error bars, respectively represent aFINM2019, aFINM without MT, and aFINM with MT. We slightly shifted the data points for aFINM without MT and aFINM with MT in time to enhance their visibility. The predicted $\mu$ values in the second panel of Figures \ref{fig:mu14} and \ref{fig:mu15} increase for the years 2009-2012 in accordance with an increase in solar activity. Using the maximum count rate in 2013 and 2014 as the peak of the distribution indicates a further increase in the predicted $\mu_{\scriptsize\textrm{eff}}$ as expected. Finally in 2017-2018 on the way toward solar minimum a decrease is evident. These variations in $\mu_{\scriptsize\textrm{eff}}$ are in accordance with previous findings \citep{Rahmanifard2019}. 

The black line in the second panel of Figures \ref{fig:mu14} and \ref{fig:mu15} shows $\mu_{\scriptsize\textrm{0}} = \mu (v_{\scriptsize\textrm{r}} = 0)$, where $v_{\scriptsize\textrm{r}}$ is the radial velocity of ISN H atoms \citep{Kowalska2018a}. $\mu_{\scriptsize\textrm{0}}$ from \citet{Kowalska2018a} clearly demonstrates the same trend as $\mu_{\scriptsize\textrm{eff}}$ derived from the IBEX observations in this study (black, red, and green circles with error bars), despite a $\sim 21\%$ difference. It is important to note that $\mu_{\scriptsize\textrm{eff}}$ is an empirical $\mu$ which may best be represented by $\mu$ values for a weighted average of $v_{\scriptsize\textrm{r}}$ in the last months of the ISN H atoms journey toward the Sun. In \citet{Rahmanifard2019}, we showed that while this $\overline{v}_{\scriptsize\textrm{r}}$ in the last 6 month of the ISN H atoms journey is $\sim 18.3$ km s\textsuperscript{–1}, $\mu_{\scriptsize\textrm{eff}}$ based on aFINM2019 is equivalent to to $\mu (v_{\scriptsize\textrm{r}} = -35  \textrm{km s\textsuperscript{–1}})$. Therefore, we expect this $\sim 21\%$ difference to be at least partially associated with absolute calibration of Ly$\alpha$ line profiles.


A linear correlation has been reported between the total Ly$\alpha$ irradiance and $\mu_{\scriptsize\textrm{0}}$ \citep{Lemaire2015}. We illustrate this correlation in Figure \ref{fig:mucor} with a blue line and blue shaded area for the uncertainty region; $\mu_{\scriptsize\textrm{0}}=(0.35 \pm 0.01) I{\scriptsize_\textrm{t}} - (0.46 \pm 0.005)$. To find $\mu_{\scriptsize\textrm{0}}$ that best describes what ISN H atoms have experienced in each season (in analogy to $\mu_{\scriptsize\textrm{eff}}$), we use the total Ly$\alpha$ irradiance averaged over two Carrington cycles before the start of each ISN H season. This approach is based on a discussion in \citet{Rahmanifard2019} where we showed the net force acting on ISN H atoms remains negligible until the last few au of their trajectory or the last two months of their journey toward the Sun. We compare this correlation (blue line) with a correlation between $\mu_{\scriptsize\textrm{eff}}$ (aFINM2019, the two cases with and without MT) and the total irradiance. The correlations for these three cases are not significantly different (shown in black, red, and green for the aFINM2019, without MT, and with MT), as expected: 

\begin{figure}[ht!]
\plotone{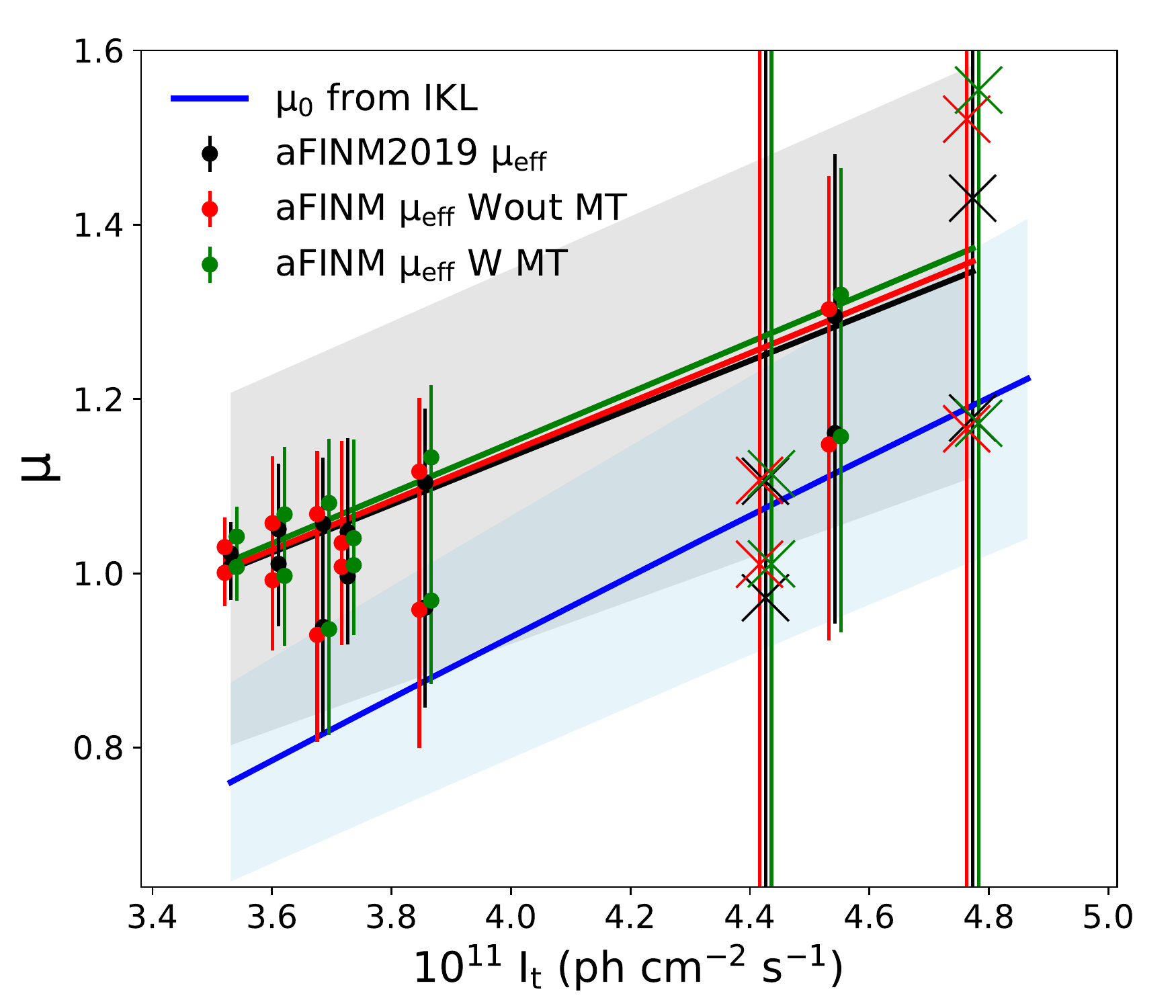}
\caption{The $\mu_{\scriptsize\textrm{eff}}$ obtained from our analysis for aFINM2019 (black), modified aFINM without MT (red), and with MT (green) circles with error bars for spin bins 14 and 15 data vs. total irradiance. The linear correlation between $\mu_{\scriptsize\textrm{0}}$ from \citet[][IKL]{Kowalska2018a} and total irradiance is shown by blue line with light blue uncertainty region. The black, red and green lines represent the linear fit to black, red and green data points respectively. The data points presented by x marks are not included in the linear fit as they are associated with the years 2013 and 2014 where a chi-square analysis is not possible. The gray uncertainty region is associated with aFINM2019 (black line). \label{fig:mucor}}
\end{figure}

\begin{equation}
\label{eq:3}
\mu{\scriptsize_\textrm{eff} ^\textrm{aFINM2019}} = (0.27 \pm 0.03) I{\scriptsize_\textrm{t}} + (0.03 \pm0.10) 
\end{equation}

\begin{equation}
\label{eq:4}
\mu{\scriptsize_\textrm{eff} ^\textrm{Wout MT}} = (0.28 \pm 0.03) I{\scriptsize_\textrm{t}} + (0.01 \pm 0.11) 
\end{equation}

\begin{equation}
\label{eq:5}
\mu{\scriptsize_\textrm{eff} ^\textrm{W MT}} = (0.29 \pm 0.03) I{\scriptsize_\textrm{t}} - (0.005 \pm 0.127)
\end{equation}

\section{Summary and Discussion} \label{sec:summary}
In this paper, we followed the methodology developed to model collisions of ISN helium atoms by \citet{Swaczyna2021, Swaczyna2023}. We used Monte Carlo simulations of test particle trajectories to simulate the transport of ISN atoms through the outer heliosheath accounting for the MT associated with the angular scattering of atoms and ions starting at 500 au with a Maxwellian distribution. We considered charge exchange and elastic collisions between ISN H atoms and protons for which we investigated two cases where these collisions occur with and without MT all the way to 100 au. We excluded collisions between hydrogen atoms. These collisions lead to thermalization of ISN hydrogen atoms and do not transfer momentum between populations.  

Using this method, as described in more detail in Section \ref{sec:heliosheath} \citep[also see][]{Swaczyna2023}, we found the ISN H atoms distribution at 100 au as a function of their entrance position and their velocity. We then used a modified version of the aFINM model \citep[Section \ref{sec:heliosphere} and ][]{Swaczyna2023} to obtain the ISN H count rates observed in the lowest energy bin of IBEX-Lo. We performed an analysis of the observed ISN H signal in the two spin bins closest to the latitudinal peak of the signal (bin 14 and 15, corresponding to $84\degr$ and $90\degr$ from the NEP).

To investigate the ISN H signal and its evolution over the solar cycle, besides the variation in the intensity of the ISN H signal, the longitudinal shift of the ISN H signal over different phases of solar activity is of key interest. These effects have been first reported by \citet{Saul2012, Saul2013}. However, due to a discrepancy in the ratio of H counts in the two lowest E-Steps between models and IBEX observations, only a few studies have addressed variations in the intensity of ISN H signal \citep{Schwadron2013, Katushkina2015} for limited IBEX observations. In \citet{Rahmanifard2019}, we used an almost entire solar cycle worth of data \citep{Galli2019} to investigate the effect of the radiation parameter on the longitudinal shift of ISN H signal. \citet{Katushkina2021} further used a combination of these two effects to investigate the radiation parameter and variations of ISN H signal through different phases of solar activity.

In this paper, we used the methodology introduced in \citet{Rahmanifard2019} to find an effective radiation parameter using the same ISN H seasons. We investigated the effect of the radiation parameter on the longitudinal shift of ISN H signal for the cases where we used aFINM2019, and modified aFINM results with and without MT. We showed the two cases with and without MT demonstrate the longitudinal shift in accordance with variations in solar activity, as does aFINM2019. Additionally, the average value for the predicted $\overline{\mu}_\textrm{eff}$ over years included in this analysis are not significantly different. 

\begin{equation}
\label{eq:6}
\overline{\mu}{\scriptsize_\textrm{eff} ^\textrm{aFINM2019}} = 1.074 \pm 0.038
\end{equation}

\begin{equation}
\label{eq:7}
\overline{\mu}{\scriptsize_\textrm{eff} ^\textrm{Wout MT}} = 1.079 \pm 0.039
\end{equation}

\begin{equation}
\label{eq:8}
\overline{\mu}{\scriptsize_\textrm{eff} ^\textrm{W MT}} = 1.087 \pm 0.040
\end{equation}

Compared to $\overline{\mu}{\scriptsize_\textrm{0}} = 0.84 \pm 0.067$  based on \citet{Kowalska2018a}, the derived $\overline{\mu}{\scriptsize_\textrm{eff}}$ values present a $\sim 21-22\%$ increase, with the case considering MT slightly larger than the case without considering the MT. Because the $\sim 21-22\%$ larger $\overline{\mu}{\scriptsize_\textrm{eff}}$ value than $\overline{\mu}{\scriptsize_\textrm{0}}$ for all the three cases investigated in this paper is in line with our previous findings from Rahmanifard (2019), these values confirm that considering charge exchange and elastic collisions and their associated angular scattering in the outer heliosheath does not introduce a significant difference in the position of the ISN H peak. Furthermore, accounting for MT in elastic collisions and charge exchange does not resolve the $\sim 21-22\%$ difference between $\overline{\mu}{\scriptsize_\textrm{0}}$ and $\overline{\mu}{\scriptsize_\textrm{eff}}$.

The IMAP-Lo instrument, part of the upcoming IMAP mission \citep[Interstellar Mapping and Acceleration Probe, ][]{McComas2018} is expected to provide higher statistics and broader and better-understood instrument response to ISN H atoms which will eventually lead to better possibilities to investigate ISN H signal. Future IMAP-Lo observations with their improved statistics, the better-defined instrument response function, and the wider range of observation season and thus vantage points thanks to the pivot platform will improve our understanding of ISN H and its fate in the heliosphere \citep{Sokol2019}. We hope that extending this kind of analysis to investigate the ISN H signal observed at different vantage points through longer observation seasons provided by IMAP-Lo will improve our understanding of the ISN H filtration and interaction in the heliospheric boundaries.

\begin{acknowledgments}
\textbf{Acknowledgments:} This work was supported by the National Aeronautics and Space Administration under grant No. 80NSSC20K0781 issued through the Outer Heliosphere Guest Investigators Program. A.G. thanks the Swiss National Science Foundation for financial support. This work was also partially funded by the IBEX mission as part of NASA’s Explorer Program (80NSSC18K0237) and IMAP mission as a part of NASA’s Solar Terrestrial Probes (STP) Program (80GSFC19C0027). J.M.S. is supported by the IBEX mission grant 80NSSC20K0719. Research at SwRI was funded by subcontract from Princeton University. 
\end{acknowledgments}


\appendix
\section{Analytic model of the ISN H distribution function at 100 au from the Sun}

The transport of ISN hydrogen atoms through the outer heliosheath is calculated using the Cartesian coordinate system in which the inflow direction coincides with the $x$-axis, the pristine interstellar magnetic field direction is within the $xy$-plane, and the $z$-axis completes the right-handed coordinate system. Because of limited statistics achievable in the Monte Carlo simulations, the distribution function reconstructed from the distribution of the velocity vectors at various entrance positions at 100 au from the Sun is limited. To provide a smooth approximation of the distribution function, we employ a two-step approach described in detail by \citet{Swaczyna2023}. 

In the first step, we calculate each population's density and bulk velocity relative to the density at the outer boundary at the inner boundary within 20 au $\times$ 20 au cells defined in the $yz$-plane. For the forward inner boundary hemisphere, the $x$-component can be computed from $y$ and $z$ as $x=(100^2-y^2-z^2 )^{1/2}$. In each cell, the density and bulk velocity components are calculated. Subsequently, the density and bulk velocity components as a function of $y$ and $z$ are approximated using a multivariate polynomial up to the degree of 4 in the following form:

\begin{equation}
\label{eq:9}
f(y,z) =  \sum_{i=0}^4 \sum_{j=0}^{4-i} a_{i,j} (\frac{y}{100 au})^i  (\frac{z}{100 au})^j
\end{equation}

In the most general form, this formula has 15 coefficients. However, because we expect a symmetry with respect to the B-V plane (i.e., $xy$-plane), the density and $x$- and $y$-components of the bulk velocity must fulfill the following condition: $f(y,-z)=f(y,z)$, which means that the coefficient $a_{i,j}$ for $j=1$, or 3, must be 0. On the other hand, for the $z$-component of the velocity, the function must satisfy $f(y,-z)=-f(y,z)$, which means that $a_{i,j}$ for $j=0,2$, or 4 must be 0, which reduces the number of non-zero coefficients in each case.

Unlike \citet{Swaczyna2023}, we do not separate the bulk velocity changes due to solar gravity. Table \ref{tab:poly} presents the found coefficients for each population's density and bulk velocity.

\begin{table}
\begin{scriptsize}
\caption{\label{tab:poly} Polynomial coefficients for densities and velocities of the ISN hydrogen populations}
\centering
\begin{ruledtabular}
\begin{tabular}{lccccccccccc}
{Population} & {Quantity}  & {MT} & \multicolumn{9}{c}{Polynomial coefficients} \\
 & (Unit) & {} & \edit1{$a_{0,0}$} & \edit1{$a_{0,2}$} & \edit1{$a_{0,4}$} & \edit1{$a_{1,0}$} & \edit1{$a_{1,2}$} & \edit1{$a_{2,0}$} & \edit1{$a_{2,2}$} & \edit1{$a_{3,0}$} & \edit1{$a_{4,0}$}  \\
\hline
Primary & $n_{pri}/n_\infty$ (1) & With & 0.224	 & –0.028	 & 0.024	 & 0.005	 & 0.016	 & –0.003	 & 0.007	 & 0.006	 & –0.004 \\
Primary & $n_{pri}/n_\infty$ (1) & Wout & 0.229	 & 0.016	 & –0.034	 & 0.006	 & 0.005	 & 0.009	 & –0.012	 & 0.007	 & –0.023 \\
Secondary & $n_{sec}/n_\infty$ (1) & With	 & 0.521	 & –0.071	 & –0.025	 & –0.044	 & 0.027	 & –0.022	 & –0.091	 & 0.041	 & –0.087 \\
Secondary & $n_{sec}/n_\infty$ (1) & Wout & 0.508	 & –0.044	 & –0.071	 & –0.001	 & –0.014	 & –0.047	 & –0.056	 & –0.022	 & –0.054 \\
Primary & $v_{pri,x}$ (km s\textsuperscript{–1}) & With & 26.46	 & 1.04	 & –0.73	 & –0.24	 & 0.06	 & 1.33	 & –1.73	 & 0.41	 & –1.24 \\
Primary & $v_{pri,x}$ (km s\textsuperscript{–1}) & Wout & 27.76	 & 0.27	 & –0.45	 & –0.07	 & –0.29	 & –0.07	 & 0.19	 & 0.13	 & –0.24 \\
Secondary & $v_{sec,x}$ (km s\textsuperscript{–1}) & With & 16.63	 & 1.73	 & 0.19	 & 0.15	 & –0.78	 & 1.41	 & –0.18	 & –0.29	 & 0.08 \\
Secondary & $v_{sec,x}$ (km s\textsuperscript{–1}) & Wout & 16.48	 & 1.63	 & 0.44	 & –0.72	 & 0.39	 & 2.07	 & –1.19	 & 0.64	 & –0.47 \\
Primary & $v_{pri,y}$ (km s\textsuperscript{–1}) & With & –0.59	 & 0.21	 & –0.34	 & –0.40	 & –0.41	 & –0.08	 & 0.66	 & –0.18 	 & 0.22 \\
Primary & $v_{pri,y}$ (km s\textsuperscript{–1}) & Wout & –0.22	 & –0.78	 & 0.96	 & –0.40	 & –0.45	 & –0.15	 & 0.34	 & –0.79	 & 0.13 \\
Secondary & $v_{sec,y}$ (km s\textsuperscript{–1}) & With	 & –2.59	 & 1.70	 & –1.10	 & 1.13	 & –1.54	 & –0.24	 & –0.27	 & –1.30	 & 1.18 \\
Secondary & $v_{sec,y}$ (km s\textsuperscript{–1}) & Wout & –2.25	 & –0.16	 & 0.75	 & 0.77	 & –0.89	 & 0.93	 & –0.47	 & –0.91	 & –0.82 \\
\hline
&  &  & \edit1{$a_{0,1}$} & \edit1{$a_{0,3}$} & \edit1{$a_{1,1}$} & \edit1{$a_{1,3}$} & \edit1{$a_{2,1}$} & \edit1{$a_{3,1}$} \\
\hline
Primary & $v_{pri,z}$ (km s\textsuperscript{–1}) & With & –0.11	& –0.11 & 1.03 & –0.98	 & –0.24	& –0.91 & & & \\
Primary & $v_{pri,z}$ (km s\textsuperscript{–1}) & Wout & –0.27	& –0.54 & 0.57 & –1.52 & –0.48	& –0.09 & & &  \\
Secondary & $v_{sec,z}$ (km s\textsuperscript{–1}) & With	 & 2.02 & –1.19	 & –0.13 & –0.65 & -1.46 & 1.09 & & & \\
Secondary & $v_{sec,z}$ (km s\textsuperscript{–1}) & Wout & 1.85 & –1.04	 & 0.21 & 0.39 & –1.18	& –0.16 & & &  \\

\end{tabular}
\end{ruledtabular}
\end{scriptsize}
\end{table}

In the second step, we find an analytic distribution function in the rotated coordinate system for each entry point at 100 au. Then, for each considered trajectory, the velocity at 100 au is rotated to the coordinate system in which the bulk flow is aligned with the $x$-axis. To avoid confusion with the $xyz$-coordinates used in the transport, we refer to these new components as parallel ($x$-axis) and perpendicular components ($\perp1$–$y$-axis, $\perp2$–$z$-axis). 
In the rotated coordinates, we assume that the distribution function is homogenous (i.e., the distribution is the same at all points on the inner boundary) and that the velocity components are independent. Therefore the distribution function can be calculated as a product of three functions applied to each component separately:

\begin{equation}
\label{eq:10}
f(v_\parallel, v_{\perp1}, v_{\perp2}) = f_\parallel(v_\parallel) f_{\perp1}(v_{\perp1}) f_{\perp2}(v_{\perp2})
\end{equation}

For the primary population, the parallel and first perpendicular ($\perp1$) components are assumed to follow a generalized asymmetric kappa distribution as derived in \citep{Swaczyna2021}. This distribution peaks at speed $u$ and is described by different temperatures and kappa indices on each side of the peak. For this asymmetric distribution function, the peak velocity $u$ differs from zero so that the resulting bulk velocity is zero in the frame co-moving with the population. The second perpendicular ($\perp2$), due to the symmetry of the problem, is described by a (symmetric) kappa distribution. The parameters of the fit functions are provided in Table \ref{tab:priparam}.

\begin{table}
\caption{\label{tab:priparam} Parameters of the primary ISN hydrogen distribution at 100 au}
\centering
\begin{ruledtabular}
\begin{tabular}{ccccccc}
{Component} & {MT} & {$u$} & { $T_1$} & {$\kappa_1$} & {$T_2$}  & {$\kappa_2$} \\
{} & {} & {(km s\textsuperscript{–1})} & {(K)} & {(1)}  & {(K)} & {(1)} \\

\hline
$v_{\parallel}$ & With & 0.21 & 7910	 & 18.7	 & 7213	 & $\infty$ \\
$v_{\parallel}$ & Wout & –0.89 & 5220	 & $\infty$	 & 7139	 & $\infty$ \\
$v_{\perp1}$	 & With & 0.03 & 8670	 & 86	          & 8515	 & $\infty$ \\
$v_{\perp1}$	 & Wout & –0.18 & 7340	 & 116	 & 7856	 & 38 \\
$v_{\perp2}$	 & With & $\equiv 0$ & 8280	 & $\infty$	 & $\equiv T_1$	 & $\equiv \kappa_1$ \\
$v_{\perp2}$	 & Wout & $\equiv 0$ & 7500	 & 77	         & $\equiv T_1$	 & $\equiv \kappa_1$ \\
\end{tabular}
\end{ruledtabular}
\end{table}

\begin{table}
\caption{\label{tab:secparam} Parameters of the secondary ISN hydrogen distribution at 100 au}
\centering
\begin{ruledtabular}
\begin{tabular}{cccccccc}
{Component} & {MT} & {$\zeta$} & {$u$} & { $T_1$} & {$\kappa_1$} & {$T_2$}  & {$\kappa_2$} \\
{} & {} & {(1)} & {(km s\textsuperscript{–1})} & {(K)} & {(1)}  & {(K)} & {(1)} \\

\hline
$v_{\parallel}(1)$ & With & 0.096	& –11.44	& 1570	& …	& …	& … \\
$v_{\parallel}(2)$ & With & 0.302	& –4.36	& 3730	& …	& …	& … \\
$v_{\parallel}(3)$ & With & 0.498	& 3.33	& 7950	& …	& …	& … \\
$v_{\parallel}(4)$ & With & 0.104	& 7.07	& 13030	& …	& …	& … \\
$v_{\parallel}(1)$ & Wout & 0.036	& –12.6	& 830	& …	& …	& … \\
$v_{\parallel}(2)$ & Wout &0.246	& –6.89	& 2960	& …	& …	& …\\
$v_{\parallel}(3)$ & Wout & 0.407	& 0.24	& 5810	& …	& …	& … \\
$v_{\parallel}(4)$ & Wout & 0.312	& 6.49	& 9670	& …	& …	& … \\
$v_{\perp1}$	 & With & $\equiv 1$ & 0.32	 & 12540	 & 18	 & 11580	 & 16 \\
$v_{\perp1}$	 & Wout & $\equiv 1$ & 0.05	 & 12290	 & 18 & 12020	 & 23 \\
$v_{\perp2}$	 & With & $\equiv 1$ & $\equiv 0$	& 9970 & 12	 & $\equiv T_1$	 & $\equiv \kappa_1$ \\
$v_{\perp2}$	 & Wout & $\equiv 1$ & $\equiv 0$	 & 9910 & 12	 & $\equiv T_1$	 & $\equiv \kappa_1$ \\
\end{tabular}
\end{ruledtabular}
\end{table}

As for helium \citep[see][]{Swaczyna2023}, the parallel component of the secondary ISN hydrogen population cannot be well described, even by the asymmetric kappa distribution. Therefore, following the procedure applied for helium, we describe this population as a superposition of 4 Maxwell distributions. However, the perpendicular components can be approximated by the kappa distributions. Table \ref{tab:secparam} shows the obtained parameters. The parameter $\zeta$ provides the relative weight of the normalized Maxwell distributions contributing to the parallel components.
Figures \ref{fig:pripop} and \ref{fig:secpop} in Section \ref{sec:heliosheath} compare the obtained histograms with the distribution functions derived above and the Maxwell approximation of each function. The distributions are non-Maxwellian. Nevertheless, the temperatures and kappa indices do not represent their thermodynamical meanings but are merely used to approximate the distribution function effectively. Finally, following the procedure outlined in \citet{Swaczyna2023}, we verified that the correlations between components are insignificant. Finally, the analytic functions are used to provide a numerical form for transport from 100 au to 1 au, as described in Section  \ref{sec:heliosphere}. 



\begin{thebibliography}{}
\expandafter\ifx\csname natexlab\endcsname\relax\def\natexlab#1{#1}\fi
\providecommand{\url}[1]{\href{#1}{#1}}
\providecommand{\dodoi}[1]{doi:~\href{http://doi.org/#1}{\nolinkurl{#1}}}
\providecommand{\doeprint}[1]{\href{http://ascl.net/#1}{\nolinkurl{http://ascl.net/#1}}}
\providecommand{\doarXiv}[1]{\href{https://arxiv.org/abs/#1}{\nolinkurl{https://arxiv.org/abs/#1}}}

\bibitem[{Bzowski \& Heerikhuisen(2020)}]{Bzowski2020}
Bzowski, M., \& Heerikhuisen, J. 2020, The Astrophysical Journal, 888, 24,
  \dodoi{10.3847/1538-4357/ab595a}

\bibitem[{Bzowski {et~al.}(2014)Bzowski, Kubiak, Hłond, Sokół,
  Banaszkiewicz, \& Witte}]{Bzowski2014}
Bzowski, M., Kubiak, M.~A., Hłond, M., {et~al.} 2014, Astronomy and
  Astrophysics, 569, A8, \dodoi{10.1051/0004-6361/201424127}

\bibitem[{Bzowski {et~al.}(2008)Bzowski, Möbius, Tarnopolski, Izmodenov, \&
  Gloeckler}]{Bzowski2008}
Bzowski, M., Möbius, E., Tarnopolski, S., Izmodenov, V.~V., \& Gloeckler, G.
  2008, Astronomy and Astrophysics, 491, 7, \dodoi{10.1051/0004-6361:20078810}

\bibitem[{Bzowski {et~al.}(2015)Bzowski, Swaczyna, Kubiak, Sokół, Fuselier,
  Galli, Heirtzler, Kucharek, Leonard, McComas, Möbius, Schwadron, \&
  Wurz}]{Bzowski2015}
Bzowski, M., Swaczyna, P., Kubiak, M.~A., {et~al.} 2015, Astrophysical Journal,
  Supplement Series, 220, \dodoi{10.1088/0067-0049/220/2/28}

\bibitem[{Bzowski {et~al.}(2019)Bzowski, Czechowski, Frisch, Fuselier, Galli,
  Grygorczuk, Heerikhuisen, Kubiak, Kucharek, McComas, Möbius, Schwadron,
  Slavin, Sokół, Swaczyna, Wurz, \& Zirnstein}]{Bzowski2019}
Bzowski, M., Czechowski, A., Frisch, P.~C., {et~al.} 2019, The Astrophysical
  Journal, 882, 60, \dodoi{10.3847/1538-4357/ab3462}

\bibitem[{Fraternale {et~al.}(2023)Fraternale, Pogorelov, \&
  Bera}]{Fraternale2023}
Fraternale, F., Pogorelov, N.~V., \& Bera, R.~K. 2023, The Astrophysical
  Journal, 946, 97, \dodoi{10.3847/1538-4357/acba10}

\bibitem[{Fuselier {et~al.}(2009)Fuselier, Bochsler, Chornay, Clark, Crew,
  Dunn, Ellis, Friedmann, Funsten, Ghielmetti, Googins, Granoff, Hamilton,
  Hanley, Heirtzler, Hertzberg, Isaac, King, Knauss, Kucharek, Kudirka, Livi,
  Lobell, Longworth, Mashburn, McComas, Möbius, Moore, Moore, Nemanich, Nolin,
  O'Neal, Piazza, Peterson, Pope, Rosmarynowski, Saul, Scherrer, Scheer,
  Schlemm, Schwadron, Tillier, Turco, Tyler, Vosbury, Wieser, Wurz, \&
  Zaffke}]{Fuselier2009}
Fuselier, S.~A., Bochsler, P., Chornay, D., {et~al.} 2009, Space Science
  Reviews, 146, 117, \dodoi{10.1007/s11214-009-9495-8}

\bibitem[{Galli {et~al.}(2019)Galli, Wurz, Rahmanifard, Möbius, Schwadron,
  Kucharek, Heirtzler, Fairchild, \& Bzowski}]{Galli2019}
Galli, A., Wurz, P., Rahmanifard, F., {et~al.} 2019, The Astrophysical Journal,
  871, \dodoi{https://doi.org/10.3847/1538-4357/aaf737}

\bibitem[{Gloeckler {et~al.}(2004)Gloeckler, Möbius, Geiss, Bzowski, Chalov,
  Fahr, McMullin, Noda, Oka, Ruciński, Skoug, Terasawa, von Steiger, Yamazaki,
  \& Zurbuchen}]{Gloeckler2004}
Gloeckler, G., Möbius, E., Geiss, J., {et~al.} 2004, Astronomy and
  Astrophysics, 426, 845, \dodoi{10.1051/0004-6361:20035768}

\bibitem[{Heerikhuisen {et~al.}(2016)Heerikhuisen, Gamayunov, Zirnstein, \&
  Pogorelov}]{Heerikhuisen2016}
Heerikhuisen, J., Gamayunov, K.~V., Zirnstein, E.~J., \& Pogorelov, N.~V. 2016,
  The Astrophysical Journal, 831, 137, \dodoi{10.3847/0004-637x/831/2/137}

\bibitem[{Heerikhuisen {et~al.}(2008)Heerikhuisen, Pogorelov, Florinski, Zank,
  \& Kharchenko}]{Heerikhuisen2009}
Heerikhuisen, J., Pogorelov, N.~V., Florinski, V., Zank, G.~P., \& Kharchenko,
  V. 2008 (ASP Conference Series), 189

\bibitem[{Katushkina {et~al.}(2021)Katushkina, Galli, Izmodenov, \&
  Alexashov}]{Katushkina2021}
Katushkina, O.~A., Galli, A., Izmodenov, V.~V., \& Alexashov, D.~B. 2021,
  Monthly Notices of the Royal Astronomical Society, 501, 1633,
  \dodoi{10.1093/mnras/staa3780}

\bibitem[{Katushkina {et~al.}(2015)Katushkina, Izmodenov, Alexashov, Schwadron,
  \& McComas}]{Katushkina2015}
Katushkina, O.~A., Izmodenov, V.~V., Alexashov, D.~B., Schwadron, N.~A., \&
  McComas, D.~J. 2015, Astrophysical Journal, Supplement Series, 220,
  \dodoi{10.1088/0067-0049/220/2/33}

\bibitem[{Kowalska-Leszczynska
  {et~al.}(2018{\natexlab{a}})Kowalska-Leszczynska, Bzowski, Sokół, \&
  Kubiak}]{Kowalska2018b}
Kowalska-Leszczynska, I., Bzowski, M., Sokół, J.~M., \& Kubiak, M.~A.
  2018{\natexlab{a}}, The Astrophysical Journal, 868, 49,
  \dodoi{10.3847/1538-4357/aae70b}

\bibitem[{Kowalska-Leszczynska
  {et~al.}(2018{\natexlab{b}})Kowalska-Leszczynska, Bzowski, Sokół, \&
  Kubiak}]{Kowalska2018a}
---. 2018{\natexlab{b}}, The Astrophysical Journal, 852, 115,
  \dodoi{10.3847/1538-4357/aa9f2a}

\bibitem[{Kowalska-Leszczynska {et~al.}(2022)Kowalska-Leszczynska, Kubiak, \&
  Bzowski}]{Kowalska2022}
Kowalska-Leszczynska, I., Kubiak, M.~A., \& Bzowski, M. 2022, The Astrophysical
  Journal, 926, 27, \dodoi{10.3847/1538-4357/ac4092}

\bibitem[{Kubiak {et~al.}(2016)Kubiak, Swaczyna, Bzowski, Sokół, Fuselier,
  Galli, Heirtzler, Kucharek, Leonard, McComas, Park, Schwadron, \&
  Wurz}]{Kubiak2016}
Kubiak, M.~A., Swaczyna, P., Bzowski, M., {et~al.} 2016, The Astrophysical
  Journal Supplement Series, 223, 13pp, \dodoi{10.3847/0067-0049/223/2/25}

\bibitem[{Lallement \& Bertin(1992)}]{Lallement1992}
Lallement, R., \& Bertin, P. 1992, Astronomy and Astrophysics, 266, 479.
\newblock \url{http://adsabs.harvard.edu/abs/1992A%26A...266..479L}

\bibitem[{Lee {et~al.}(2012)Lee, Kucharek, Möbius, Wu, Bzowski, \&
  McComas}]{Lee2012}
Lee, M.~A., Kucharek, H., Möbius, E., {et~al.} 2012, Astrophysical Journal,
  Supplement Series, 198, \dodoi{10.1088/0067-0049/198/2/10}

\bibitem[{Lee {et~al.}(2015)Lee, Möbius, \& Leonard}]{Lee2015}
Lee, M.~A., Möbius, E., \& Leonard, T.~W. 2015, The Astrophysical Journal
  Supplement Series, 220, 23, \dodoi{10.1088/0067-0049/220/2/23}

\bibitem[{Lemaire {et~al.}(2015)Lemaire, Vial, Curdt, Schühle, \&
  Wilhelm}]{Lemaire2015}
Lemaire, P., Vial, J.~C., Curdt, W., Schühle, U., \& Wilhelm, K. 2015,
  Astronomy and Astrophysics, 581, A26, \dodoi{10.1051/0004-6361/201526059}

\bibitem[{Linsky {et~al.}(1993)Linsky, Brown, Gayley, \& Diplas}]{Linsky1993}
Linsky, J.~L., Brown, A., Gayley, K., \& Diplas, A. 1993, The Astrophysical
  Journal, 402, 694, \dodoi{10.1086/172170}

\bibitem[{McComas {et~al.}(2009)McComas, Allegrini, Bochsler, Bzowski,
  Christian, Crew, Demajistre, Fahr, Fichtner, Frisch, Funsten, Fuselier,
  Gloeckler, Gruntman, Heerikhuisen, Izmodenov, Janzen, Knappenberger,
  Krimigis, Kucharek, Lee, Livadiotis, Livi, MacDowall, Mitchell, Möbius,
  Moore, Pogorelov, Reisenfeld, Roelof, Saul, Schwadron, Valek, Vanderspek,
  Wurz, \& Zank}]{McComas2009}
McComas, D.~J., Allegrini, F., Bochsler, P., {et~al.} 2009, Science, 326, 959,
  \dodoi{10.1126/science.1180906}

\bibitem[{McComas {et~al.}(2012)McComas, Alexashov, Bzowski, Fahr,
  Heerikhuisen, Izmodenov, Lee, Möbius, Pogorelov, Schwadron, \&
  Zank}]{McComas2012}
McComas, D.~J., Alexashov, D.~B., Bzowski, M., {et~al.} 2012, Science, 336,
  1291, \dodoi{10.1126/science.1221054}

\bibitem[{McComas {et~al.}(2015)McComas, Bzowski, Fuselier, Frisch, Galli,
  Izmodenov, Katushkina, Kubiak, Lee, Leonard, Möbius, Park, Schwadron,
  Sokół, Swaczyna, Wood, \& Wurz}]{McComas2015}
McComas, D.~J., Bzowski, M., Fuselier, S.~A., {et~al.} 2015, Astrophysical
  Journal, Supplement Series, 220, \dodoi{10.1088/0067-0049/220/2/22}

\bibitem[{McComas {et~al.}(2018)McComas, Christian, Schwadron, Fox, Westlake,
  Allegrini, Baker, Biesecker, Bzowski, Clark, Cohen, Cohen, Dayeh, Decker,
  de~Nolfo, Desai, Ebert, Elliott, Fahr, Frisch, Funsten, Fuselier, Galli,
  Galvin, Giacalone, Gkioulidou, Guo, Horanyi, Isenberg, Janzen, Kistler,
  Korreck, Kubiak, Kucharek, Larsen, Leske, Lugaz, Luhmann, Matthaeus,
  Mitchell, Möbius, Ogasawara, Reisenfeld, Richardson, Russell, Sokół,
  Spence, Skoug, Sternovsky, Swaczyna, Szalay, Tokumaru, Wiedenbeck, Wurz,
  Zank, \& Zirnstein}]{McComas2018}
McComas, D.~J., Christian, E.~R., Schwadron, N.~A., {et~al.} 2018, Space
  Science Reviews, 214, 116, \dodoi{10.1007/s11214-018-0550-1}

\bibitem[{Möbius {et~al.}(2004)Möbius, Bzowski, Chalov, Fahr, Gloeckler,
  Izmodenov, Kallenbach, Lallement, McMullin, Noda, Oka, Pauluhn, Raymond,
  Ruciński, Skoug, Terasawa, Thompson, Vallerga, von Steiger, \&
  Witte}]{Mobius2004}
Möbius, E., Bzowski, M., Chalov, S., {et~al.} 2004, Astronomy and
  Astrophysics, 426, 897, \dodoi{10.1051/0004-6361:20035834}

\bibitem[{Möbius {et~al.}(2009)Möbius, Bochsler, Bzowski, Crew, Funsten,
  Fuselier, Ghielmetti, Heirtzler, Izmodenov, Kubiak, Kucharek, Lee, Leonard,
  McComas, Petersen, Saul, Scheer, Schwadron, Witte, \& Wurz}]{Mobius2009}
Möbius, E., Bochsler, P., Bzowski, M., {et~al.} 2009, Science, 326, 969,
  \dodoi{10.1126/science.1180971}

\bibitem[{Ovchinnikov {et~al.}(2017)Ovchinnikov, Kamyshkov, Zaman, \&
  Schultz}]{Ovchinnikov2017}
Ovchinnikov, S.~Y., Kamyshkov, Y., Zaman, T., \& Schultz, D.~R. 2017, Journal
  of Physics B: Atomic, Molecular and Optical Physics, 50,
  \dodoi{10.1088/1361-6455/aa64ac}

\bibitem[{Rahmanifard {et~al.}(2019)Rahmanifard, Möbius, Schwadron, Galli,
  Richards, Kucharek, Sokół, Heirtzler, Lee, Bzowski, Kowalska-Leszczynska,
  Kubiak, Wurz, Fuselier, \& McComas}]{Rahmanifard2019}
Rahmanifard, F., Möbius, E., Schwadron, N., {et~al.} 2019, Astrophysical
  Journal, 887, \dodoi{10.3847/1538-4357/ab58ce}

\bibitem[{Saul {et~al.}(2012)Saul, Wurz, Rodriguez, Scheer, Möbius, Schwadron,
  Kucharek, Leonard, Bzowski, Fuselier, Crew, \& McComas}]{Saul2012}
Saul, L., Wurz, P., Rodriguez, D., {et~al.} 2012, The Astrophysical Journal
  Supplement Series, 198, 14, \dodoi{10.1088/0067-0049/198/2/14}

\bibitem[{Saul {et~al.}(2013)Saul, Bzowski, Fuselier, Kubiak, McComas, Möbius,
  Sokół, Rodríguez, Scheer, \& Wurz}]{Saul2013}
Saul, L., Bzowski, M., Fuselier, S., {et~al.} 2013, Astrophysical Journal, 767,
  1, \dodoi{10.1088/0004-637X/767/2/130}

\bibitem[{Schultz {et~al.}(2016)Schultz, Ovchinnikov, Stancil, \&
  Zaman}]{Schultz2016}
Schultz, D.~R., Ovchinnikov, S.~Y., Stancil, P.~C., \& Zaman, T. 2016, Journal
  of Physics B: Atomic, Molecular and Optical Physics, 49,
  \dodoi{10.1088/0953-4075/49/8/084004}

\bibitem[{Schwadron {et~al.}(2013)Schwadron, Moebius, Kucharek, Lee, French,
  Saul, Wurz, Bzowski, Fuselier, Livadiotis, Mccomas, Frisch, Gruntman, \&
  Mueller}]{Schwadron2013}
Schwadron, N.~A., Moebius, E., Kucharek, H., {et~al.} 2013, The Astrophysical
  Journal, 775, \dodoi{10.1088/0004-637X/775/2/86}

\bibitem[{Schwadron {et~al.}(2015)Schwadron, Möbius, Leonard, Fuselier,
  McComas, Heirtzler, Kucharek, Rahmanifard, Bzowski, Kubiak, Sokół,
  Swaczyna, \& Frisch}]{Schwadron2015}
Schwadron, N.~A., Möbius, E., Leonard, T., {et~al.} 2015, Astrophysical
  Journal, Supplement Series, 220, \dodoi{10.1088/0067-0049/220/2/25}

\bibitem[{Schwadron {et~al.}(2016)Schwadron, Möbius, McComas, Bochsler,
  Bzowski, Fuselier, Livadiotis, Frisch, Müller, Heirtzler, Kucharek, \&
  Lee}]{Schwadron2016}
Schwadron, N.~A., Möbius, E., McComas, D.~J., {et~al.} 2016, The Astrophysical
  Journal, 828, 81, \dodoi{10.3847/0004-637X/828/2/81}

\bibitem[{Schwadron {et~al.}(2022)Schwadron, Möbius, McComas, Bower, Bower,
  Bzowski, Fuselier, Heirtzler, Kubiak, Lee, Rahmanifard, Sokół, Swaczyna, \&
  Winslow}]{Schwadron2022}
---. 2022, The Astrophysical Journal Supplement Series, 258, 7,
  \dodoi{10.3847/1538-4365/ac2fa9}

\bibitem[{Sokół \& Bzowski(2014)}]{Sokol2014}
Sokół, J.~M., \& Bzowski, M. 2014, arXiv.org.
\newblock \url{https://arxiv.org/abs/1411.4826}

\bibitem[{Sokół {et~al.}(2013)Sokół, Bzowski, Tokumaru, Fujiki, \&
  Mccomas}]{Sokol2013}
Sokół, J.~M., Bzowski, M., Tokumaru, M., Fujiki, K., \& Mccomas, D.~J. 2013,
  Sol Phys, 285, 167, \dodoi{10.1007/s11207-012-9993-9}

\bibitem[{Sokół {et~al.}(2019)Sokół, Kubiak, Bzowski, Möbius, \&
  Schwadron}]{Sokol2019}
Sokół, J.~M., Kubiak, M.~A., Bzowski, M., Möbius, E., \& Schwadron, N.~A.
  2019, The Astrophysical Journal Supplement Series, 245, 28,
  \dodoi{10.3847/1538-4365/ab50bc}

\bibitem[{Sokół {et~al.}(2020)Sokół, McComas, Bzowski, \&
  Tokumaru}]{Sokol2020}
Sokół, J.~M., McComas, D.~J., Bzowski, M., \& Tokumaru, M. 2020, The
  Astrophysical Journal, 897, 179, \dodoi{10.3847/1538-4357/ab99a4}

\bibitem[{Swaczyna {et~al.}(2019)Swaczyna, McComas, Zirnstein, \&
  Heerikhuisen}]{Swaczyna2019}
Swaczyna, P., McComas, D.~J., Zirnstein, E.~J., \& Heerikhuisen, J. 2019, The
  Astrophysical Journal, 887, 223, \dodoi{10.3847/1538-4357/ab5440}

\bibitem[{Swaczyna {et~al.}(2023{\natexlab{a}})Swaczyna, Rahmanifard,
  Zirnstein, \& Heerikhuisen}]{Swaczyna2023}
Swaczyna, P., Rahmanifard, F., Zirnstein, E.~J., \& Heerikhuisen, J.
  2023{\natexlab{a}}, The Astrophysical Journal, 943, 74,
  \dodoi{10.3847/1538-4357/acaa36}

\bibitem[{Swaczyna {et~al.}(2021)Swaczyna, Rahmanifard, Zirnstein, McComas, \&
  Heerikhuisen}]{Swaczyna2021}
Swaczyna, P., Rahmanifard, F., Zirnstein, E.~J., McComas, D.~J., \&
  Heerikhuisen, J. 2021, The Astrophysical Journal Letters, 911, L36,
  \dodoi{10.3847/2041-8213/abf436}

\bibitem[{Swaczyna {et~al.}(2022)Swaczyna, Kubiak, Bzowski, Bower, Fuselier,
  Galli, Heirtzler, McComas, Möbius, Rahmanifard, \& Schwadron}]{Swaczyna2022}
Swaczyna, P., Kubiak, M.~A., Bzowski, M., {et~al.} 2022, The Astrophysical
  Journal Supplement Series, 259, 42, \dodoi{10.3847/1538-4365/ac4bde}

\bibitem[{Swaczyna {et~al.}(2023{\natexlab{b}})Swaczyna, Bzowski, Fuselier,
  Galli, Heerikhuisen, Kubiak, McComas, Möbius, Rahmanifard, \&
  Schwadron}]{Swaczyna2023_HeResponse}
Swaczyna, P., Bzowski, M., Fuselier, S.~A., {et~al.} 2023{\natexlab{b}}, The
  Astrophysical Journal Supplement Series, 266, 2,
  \dodoi{10.3847/1538-4365/acc397}

\bibitem[{Vallerga {et~al.}(2004)Vallerga, Lallement, Lemoine, Dalaudier, \&
  McMullin}]{Vallerga2004}
Vallerga, J., Lallement, R., Lemoine, M., Dalaudier, F., \& McMullin, D. 2004,
  Astronomy and Astrophysics, 426, 855, \dodoi{10.1051/0004-6361:20035887}

\bibitem[{Witte(2004)}]{Witte2004}
Witte, M. 2004, Astronomy and Astrophysics, 426, 835,
  \dodoi{10.1051/0004-6361:20035956}

\bibitem[{Wood {et~al.}(2015)Wood, Müller, Bzowski, Sokół, Möbius, Witte,
  \& McComas}]{Wood2015}
Wood, B.~E., Müller, H.~R., Bzowski, M., {et~al.} 2015, Astrophysical Journal,
  Supplement Series, 220, \dodoi{10.1088/0067-0049/220/2/31}

\bibitem[{Wurz {et~al.}(2008)Wurz, Saul, Scheer, Möbius, Kucharek, \&
  Fuselier}]{Wurz2008}
Wurz, P., Saul, L., Scheer, J.~A., {et~al.} 2008, Journal of Applied Physics,
  103, 054904, \dodoi{10.1063/1.2842398}

\bibitem[{Wurz {et~al.}(1997)Wurz, Schletti, \& Aellig}]{Wurz1997}
Wurz, P., Schletti, R., \& Aellig, M.~R. 1997, Surface Science, 373, 56.
\newblock \url{http://archive.space.unibe.ch/staff/wurz/Wurz1997.pdf}

\bibitem[{Zank {et~al.}(2013)Zank, Heerikhuisen, Wood, Pogorelov, Zirnstein, \&
  McComas}]{Zank2013}
Zank, G.~P., Heerikhuisen, J., Wood, B.~E., {et~al.} 2013, Astrophysical
  Journal, 763, \dodoi{10.1088/0004-637X/763/1/20}

\bibitem[{Zirnstein {et~al.}(2016)Zirnstein, Heerikhuisen, Funsten, Livadiotis,
  McComas, \& Pogorelov}]{Zirnstein2016}
Zirnstein, E.~J., Heerikhuisen, J., Funsten, H.~O., {et~al.} 2016, The
  Astrophysical Journal, 818, L18, \dodoi{10.3847/2041-8205/818/1/l18}

\bibitem[{Zirnstein {et~al.}(2013)Zirnstein, Heerikhuisen, McComas, \&
  Schwadron}]{Zirnstein2013}
Zirnstein, E.~J., Heerikhuisen, J., McComas, D.~J., \& Schwadron, N.~A. 2013,
  Astrophysical Journal, 778, \dodoi{10.1088/0004-637X/778/2/112}

\end{thebibliography}

\bibliographystyle{aasjournal}



\end{document}